\documentclass[11pt,a4paper,DIV11]{scrartcl}
\usepackage[utf8]{inputenc}
\usepackage[T1]{fontenc}
\usepackage{lmodern}
\usepackage[british]{babel}
\usepackage{amsmath}
\usepackage{amssymb}
\usepackage{amsfonts}
\usepackage{color}
\usepackage{float}
\usepackage{hyperref}
\usepackage{graphicx}
\DeclareMathOperator{\Di}{D_w}
\newcommand{\tr}{\ensuremath{\mathrm{tr}}}
\newcommand{\I}{\ensuremath{\mathrm{i}}}
\newcommand{\arxiv}[1]{arXiv:\,\href{http://arxiv.org/abs/#1}{{\tt #1}}}
\newcommand{\aetap}{\text{a--}\eta'}
\newcommand{\api}{\text{a--}\pi}
\newcommand{\afn}{\text{a--}f_0}
\newcommand{\sapi}{\text{a}\mbox{-}\pi}
\newcommand{\glg}{\tilde{g}g}
\begin{document}
\title{
  {\vspace{-2cm}\normalsize
   \hfill\parbox[b][30mm][t]{35mm}{\textmd{MS-TP-13-07\\DESY 13-065}}}\\[-10mm]
Towards the spectrum of low-lying particles in supersymmetric
Yang-Mills theory
\vspace*{3mm}}
\author{G.~Bergner\\
\textit{\large Universit\"at Frankfurt, Institut f\"ur Theoretische Physik}\\
\textit{\large Max-von-Laue-Str.~1, D-60438 Frankfurt am Main, Germany}\\ 
\textit{\large E-mail: g.bergner@uni-muenster.de}\\[8mm]
I.~Montvay\\
\textit{\large Deutsches Elektronen-Synchrotron DESY}\\
\textit{\large Notkestr. 85, D-22603 Hamburg, Germany}\\[8mm]
G.~M\"unster, U.\,D.~\"Ozugurel, D.~Sandbrink\\
\textit{\large Universit\"at M\"unster, Institut f\"ur Theoretische Physik}\\
\textit{\large Wilhelm-Klemm-Str.~9, D-48149 M\"unster, Germany}
\vspace*{5mm}}

\date{October 15, 2013}

\maketitle

\begin{abstract}
The non-perturbative properties of supersymmetric theories are of interest
for elementary particle physics beyond the Standard Model. Numerical
simulations of these theories are associated with theoretical and technical
challenges. The minimal supersymmetric model containing gauge fields is the
$\mathcal{N}=1$ supersymmetric Yang-Mills theory. We present the results of
our investigations of the masses of the lightest particles of this model on
a lattice. The central question is, whether a continuum limit exists with
unbroken supersymmetry. In this case the bound states would form
mass-degenerate supermultiplets. We have obtained the masses of the
gluino-glue particle, mesonic states, and the scalar glueball at a fine
lattice spacing. The statistical accuracy as well as the control of finite
size effects and lattice artefacts are significantly better than in all
previous investigations. Taking the statistical and systematic uncertainties
into account, the masses of the fermionic and bosonic states in our present
calculations are consistent with the formation of degenerate
supermultiplets, indicating that in the continuum limit there is no
spontaneous supersymmetry breaking. This new finding is in contrast to
previous results.
\end{abstract}
%%%%%%%%%%%%%%%%%%%%%%%%%%%%%%%%%%%%%%%%%%%%%%%%%%%%%%%%%%%%%%%%%%%%%%%%
\newpage
\section{Introduction}

Supersymmetry plays a central role in theoretical models for elementary
particle physics beyond the Standard Model. Therefore it is important to
gain knowledge about the properties of supersymmetric theories. Much of what
is known about supersymmetric models is based on tree-level considerations
or comes from perturbation theory. However, various important
characteristics, like the existence of mass-degenerate supermultiplets of
particles, are of a non-perturbative nature. It is hence desirable to study
them by means of non-perturbative methods like numerical lattice
simulations.

The $\mathcal{N}=1$ supersymmetric Yang-Mills theory (SYM) is the minimal
supersymmetric extension of a non-Abelian gauge theory. It is a theory of
gluons, described by gauge theory, and their superpartners, the gluinos,
which are spin 1/2 Majorana fermions in the adjoint representation of the
gauge group. If supersymmetry is not broken spontaneously, the bound states
described by SYM should form supermultiplets, whose members have identical
masses.

In previous investigations of SYM by Monte Carlo simulations on space-time
lattices the expected degeneracy of the fermionic and bosonic masses was not
observed \cite{Demmouche:2010sf,Bergner:2011wf}. The question, whether this
is an indication for supersymmetry breaking or not, demands a close
examination of possible systematic errors. This is the motivation for our
present work.

%%%%%%%%%%%%%%%%%%%%%%%%%%%%%%%%%%%%%%%%%%%%%%%%%%%%%%%%%%%%%%%%%%%%%%%%
\section{The model}

The on-shell Lagrangian of SYM in Minkowski space is
\begin{equation}
\mathcal{L}=\tr\left[-\frac{1}{4}
F_{\mu\nu}F^{\mu\nu}+\frac{\I}{2}\bar{\lambda}\gamma^\mu
D_\mu\lambda{-\frac{m_g}{2}\bar{\lambda}\lambda} \right ] \,,
\end{equation}
where $F_{\mu\nu}$ is the non-Abelian field strength formed out of the gluon
fields $A_{\mu}(x)$, $\lambda(x)$ is the gluino field, and $D_\mu$ denotes
the gauge covariant derivative in the adjoint representation. The
supersymmetry of the theory is broken softly by the gluino mass term.

In several aspects the model is similar to QCD \cite{Amati:1988ft}. The
importance of related theories in extensions of the Standard Model and the
connection to QCD \cite{Armoni:2009zz} is a main motivation for the recent
interest in these models. SYM is asymptotically free and is assumed to show
confinement. Thus gluons and gluinos are not particle states in the physical
Hilbert space, which instead contains bound states of gluons and gluinos.
The determination of the masses of bound states is a non-perturbative
problem. Previous work of our group has been dedicated to the calculation of
the low-lying masses of bound states in SYM by means of numerical
simulations on a space-time lattice \cite{Demmouche:2010sf,Bergner:2011wf}.
The present article is a continuation of our preparatory work, which should
be referred to for more details and references \cite{Bergner:2012rv}.

On the lattice, supersymmetry is generically broken \cite{Bergner:2009vg}.
In SYM a fine-tuning of the bare gluino mass parameter in the continuum
limit is enough to approach the symmetries of the continuum theory
\cite{Curci:1986sm,Suzuki:2012pc}. These symmetries include (spontaneously
broken) chiral symmetry and supersymmetry. The theoretical prediction of the
existence of a supersymmetric chiral continuum limit needs to be confronted
with the numerical lattice simulations.

A necessary condition for the restoration of supersymmetry in the continuum
limit is the degeneracy of fermionic and bosonic masses. From low energy
effective theories, predictions have been made for two low-lying
supermultiplets \cite{Veneziano:1982ah,Farrar:1997fn}. Each multiplet
consists of a scalar, a pseudoscalar, and a fermionic spin 1/2 particle. The
lighter multiplet consists of a $0^{++}$ glueball, a $0^{-+}$ glueball, and
a gluino-glue state. The gluino-glue is an exotic spin 1/2 Majorana fermion,
which can be created by the operator
\begin{equation}
\label{eq:gluinocont}
\tilde{O}_{g\tilde{g}}=
\sum_{\mu\nu} \sigma_{\mu\nu} \tr \left[ F^{\mu\nu} \lambda\right],
\end{equation}
with $\sigma_{\mu\nu}=\frac{1}{2} \left[ \gamma_\mu,\gamma_\nu \right]$. The
heavier multiplet is built from the scalar meson $\afn$, represented by
$\bar{\lambda}\lambda$, the pseudo-scalar meson $\aetap$, represented by
$\bar{\lambda} \gamma_5 \lambda$, and a gluino-glue state.

In our previous work \cite{Demmouche:2010sf,Bergner:2011wf} the gluino-glue
appeared heavier than its lightest possible superpartners. In that work,
however, the masses were obtained at a fixed lattice spacing, and the
influence of finite volume effects was not known with sufficient accuracy.
In order to obtain results relevant for the presumed supersymmetric
continuum limit, the chiral limit, the continuum limit, and the infinite
volume limit have to be extrapolated from the simulations.

In our preparatory work \cite{Bergner:2012rv} we have made a detailed
analysis of finite volume effects, and observed supersymmetry breaking
effects due to the finite spatial extent of the lattice. The results of
these calculations allow to estimate the lattice sizes necessary for
neglecting them. 

In the present article we conclude our investigations of the finer lattice
spacing, taking into account also different levels of stout smearing. The
masses of the gluino-glue particle, the $\aetap$, the $\afn$ meson, and the
scalar glueball are obtained with the high statistics that turned out to be
necessary, and extrapolations towards vanishing gluino mass are made. The
statistical accuracy as well as the control of finite size effects and
lattice artefacts is better than in all the previous investigations.

%%%%%%%%%%%%%%%%%%%%%%%%%%%%%%%%%%%%%%%%%%%%%%%%%%%%%%%%%%%%%%%%%%%%%%%%
\section{Numerical simulations}

In our numerical simulations of SYM, the Euclidean version of the model with
gauge group SU(2) is formulated on a space-time lattice with an action
proposed by \cite{Curci:1986sm}. The gauge field dynamics is defined by the
tree-level Symanzik improved plaquette action. The inverse bare gauge
coupling in the current simulations was $\beta=1.75$. The results will be
compared to the previous ones at $\beta=1.6$
\cite{Demmouche:2010sf,Bergner:2011wf}. The gluinos are described by Wilson
fermions in the adjoint representation. The Wilson-Dirac operator
\begin{multline}
(\Di)_{x,a,\alpha;y,b,\beta}\\
 = \delta_{xy}\delta_{ab}\delta_{\alpha\beta}
 -\kappa\sum_{\mu=1}^{4}
  \left[(1-\gamma_\mu)_{\alpha\beta}(V_\mu(x))_{ab} \delta_{x+\mu,y}
  +(1+\gamma_\mu)_{\alpha\beta}(V^\dag_\mu(x-\mu))_{ab} 
   \delta_{x-\mu,y}\right]
\end{multline}
contains stout smeared \cite{Morningstar:2003gk} gauge links $V_{\mu}(x)$ in
the adjoint representation. The hopping parameter $\kappa$ is related to the
bare gluino mass via $\kappa=1/(2m_g+8)$. The recovery of both supersymmetry
and chiral U(1)$_R$ symmetry in the continuum limit requires to tune the
hopping parameter to the point $\kappa_c(\beta)$, where the renormalised
gluino mass vanishes \cite{Curci:1986sm,Suzuki:2012pc}. In practice, this is
achieved by monitoring the mass of the unphysical adjoint pion ($\api$),
which is the pion in the corresponding theory with two Majorana fermions
in the adjoint representation. The correlator of this particle is the   
connected contribution of the $\aetap$ correlator. 

The $\api$ is not a physical particle in SYM, which only contains one
Majorana fermion. However, it can be defined in a partially quenched setup,
in which the model is supplemented by a second species of gluinos and the
corresponding bosonic ghost gluinos \cite{Stuewe}, in the same way as for
one-flavour QCD \cite{Farchioni:2007dw}. On the basis of arguments involving
the OZI-approximation of SYM \cite{Veneziano:1982ah}, the adjoint pion mass
is expected to vanish for a massless gluino and the behaviour $m^2_{\api}
\propto m_g$ can be assumed for light gluinos. This behaviour is indeed
found in partially quenched chiral perturbation theory. Alternatively, the
gluino mass can be obtained from the lattice SUSY Ward-Identities (WIs) as
discussed in \cite{Farchioni:2001wx}. Both the gluino mass from SUSY WIs and
the adjoint pion mass have been studied numerically
in~\cite{Demmouche:2010sf}. The results show that both the WIs and adjoint
pion mass methods give consistent estimates of the critical hopping
parameter $\kappa_{c}$, and that $m^2_{\api}$ is proportioal to $m_g$. The
$\api$, however, yields a more precise signal for the tuning than the
supersymmetric Ward identities.

In our simulations the configurations have been obtained by updating with a
two-step polynomial hybrid Monte Carlo (PHMC) algorithm
\cite{Montvay:2005tj,Demmouche:2010sf}. Near $\kappa_c$ the occurrence of
low eigenvalues of the Hermitian Wilson-Dirac operator makes it necessary to
introduce correction factors to the polynomial approximation in the PHMC
algorithm. These have been obtained, when necessary, from the correct
fermionic contribution of the lowest eigenvalues.

In contrast to the theory in the continuum, the lattice theory has a (mild)
sign problem. The Pfaffian obtained by the integration of the Majorana
fermions can sometimes have a negative sign \cite{Bergner:2011zp}. When
necessary, we included this sign in a reweighting. To reduce the statistical
errors we have chosen the parameters of our present simulations such that
the reweighting with correction factors and Pfaffian signs is not relevant
for the results. The number of reweighting factors smaller than $0.98$ is
around $3\%$ at $\kappa=0.1495$ and around $1\%$ at $\kappa=0.1494$
($l_s=1$; $32^3\times 64$). In all other runs the reweighting is not taken
into account because the number of negative Pfaffians is always below $1\%$.

On the basis of our investigation of finite volume effects
\cite{Bergner:2012rv}, the lattice sizes have been chosen such that finite
volume effects can be neglected in comparison with statistical errors.

For a comparison of dimensionful quantities we use the Sommer parameter
$r_0$, obtained from the static quark potential. For illustration, QCD units
are being used by setting the Sommer parameter to $r_0=0.5 \,\mathrm{fm}$.
The value of $r_0$ is the result of an extrapolation to $\kappa_c$ of the
data at $\kappa > \kappa_c$. In QCD units the lattice spacing is $a=0.058
\,\mathrm{fm}$ at $\beta=1.75$, and $a=0.088 \,\mathrm{fm}$ at $\beta=1.6$
for the runs with one level of stout smearing. A convenient substitute for
the gluino mass is the squared adjoint pion mass in physical units, $(r_0
m_{\sapi})^2$.

%%%%%%%%%%%%%%%%%%%%%%%%%%%%%%%%%%%%%%%%%%%%%%%%%%%%%%%%%%%%%%%%%%%%%%%%
\section{Bound state masses}

We have investigated the masses of the gluino-glue particle, the $\aetap$
and the $\afn$ mesons, and the scalar glueball. The masses of the bound
states are obtained from fits to the corresponding correlation functions. In
case of the $\api$ and the gluino-glue the correlation functions yield
rather good fits for a number of fit intervals. By means of a histogram
method \cite{Baron:2010th} reliable mass estimates for these particles could
be obtained. In case of the mesons $\aetap$ and $\afn$ the mass estimate is
based on a single optimally selected interval in Euclidean time. This
interval is fixed to the region where a plateau of the effective mass is
observed. A second estimate is taken from a fit interval of the same length,
but shifted by one positive unit in Euclidean time. The difference of these
two masses provides an estimate of the systematic error of the procedure.
Further details of the histogram method and the fit procedure are explained
in \cite{Bergner:2012rv}. The fit procedure for the glueball is detailed
below. The results for the masses are collected in Table \ref{tab:mass}.

%%%%%%%%%%%%%%%%%%%%%%%%%%%%%%%%%%%%%%%%%
\begin{table}[htb]
\begin{center}
\begin{tabular}{|l|c|c|c|l|l|l|l|l|}
\hline
\hspace{5mm}$\kappa$ & $l_s$ & $L$ & $T$ & \hspace{1mm}$r_0 m_{\api}$
 & \hspace{5mm}$r_0 m_{\aetap}$ & \hspace{5mm}$r_0 m_{\afn}$
 & \hspace{2mm}$r_0 m_{\glg}$ & \hspace{5mm}$r_0 m_{0^{++}}$\\
\hline
$0.1490$  & 1 & 24 & 48 & $2.145(51)$ & $3.39(21)(10)$   & $3.60(36)(39)$   & $3.65(27)$  & $4.39(26)(25)$\\
$0.1490$  & 1 & 32 & 64 & $2.151(46)$ & $3.14(16)(12)$   & $5.34(72)(97)$   & $3.44(25)$  & $4.26(43)(50)$\\
$0.1492$  & 1 & 24 & 48 & $1.829(43)$ & $2.92(17)(21)$   & $2.53(31)(51)$   & $3.36(16)$  & $3.78(26)(31)$\\
$0.1492$  & 1 & 32 & 64 & $1.835(41)$ & $2.84(15)(21)$   & $3.58(43)(44)$   & $3.25(19)$  & $3.84(32)(35)$\\
$0.14925$ & 1 & 24 & 48 & $1.710(49)$ & $2.78(20)(21)$   & $3.55(65)(17)$   & $2.97(16)$  & $2.72(29)(41)$\\
$0.1493$  & 1 & 24 & 48 & $1.51(11)$  & $2.82(25)(14)$   & $2.24(49)(28)$   & $2.86(23)$  & $3.16(30)(88)$\\
$0.1494$  & 1 & 32 & 64 & $1.447(42)$ & $2.70(23)(22)$   & $3.94(47)(35)$   & $3.17(19)$  & $4.36(38)(29)$\\
$0.1495$  & 1 & 32 & 64 & $1.167(44)$ & $2.52(34)(21)$   & $1.50(80)(22)$   & $2.85(21)$  & $4.15(45)(63)$\\
\hline
$0.1350$  & 3 & 24 & 64 & $3.574(38)$ & $4.113(91)(16)$  & $4.58(51)(30)$   & $4.77(15)$  & $4.609(320)(77)$\\
$0.1355$  & 3 & 24 & 64 & $3.117(35)$ & $3.779(97)(170)$ & $4.18(40)(26)$   & $4.21(27)$  & $4.27(28)(13)$\\
$0.1360$  & 3 & 24 & 64 & $2.584(32)$ & $3.319(88)(160)$ & $3.47(30)(42)$   & $4.10(18)$  & $3.32(22)(48)$\\
$0.1365$  & 3 & 24 & 64 & $1.946(24)$ & $2.92(13)(17)$   & $3.488(270)(47)$ & $3.47(14)$  & $3.92(25)(16)$\\
$0.1368$  & 3 & 24 & 64 & $1.445(42)$ & $2.30(24)(27)$   & $3.08(23)(16)$   & $3.143(91)$ & $3.49(21)(37)$\\
$0.1368$  & 3 & 32 & 64 & $1.475(21)$ & $2.36(17)(22)$   & $2.76(42)(33)$   & $3.09(10)$  & $2.31(21)(58)$\\
\hline
\end{tabular}
\caption{The masses of the relevant bound states in units of the Sommer
scale $r_0$. $l_s$ is the level of stout smearing, and $L,T$ are the spatial
and temporal extents of the lattice. The value of $r_0/a$ being used is
obtained from the extrapolation to the chiral limit. Its value is
$r_0/a=9.02(18)$ for $l_s=1$, and $r_0/a=8.663(81)$ for $l_s=3$. For the
mesons $\aetap$ and $\afn$ and for the glueball $0^{++}$ the indicated error
is given as (statistical error)(systematic error of the plateau estimation).
The $\api$ and gluino-glue ($\glg$) mass is obtained with the histogram
method and already includes an estimate of the systematic error.}
\label{tab:mass}
\end{center}
\end{table}
%%%%%%%%%%%%%%%%%%%%%%%%%%%%%%%%%%%%%%%%%

The extrapolation to the chiral limit is obtained from a linear fit of the
masses as a function of the squared adjoint pion mass. Up to lattice
artifacts and finite size effects this limit coincides with the
supersymmetric limit, where the particle masses should be grouped in the
predicted multiplets.

For the lattices with spatial extent $L=24$ our statistics is much higher
than for $L=32$. An extent of $L=24$ is also sufficiently large in view of
finite size effects. Therefore we have decided to include only the $L=24$
data in the extrapolations. Adding the $L=32$ data would lead to small
changes only.

\begin{figure}[tb]
\begin{center}
\includegraphics{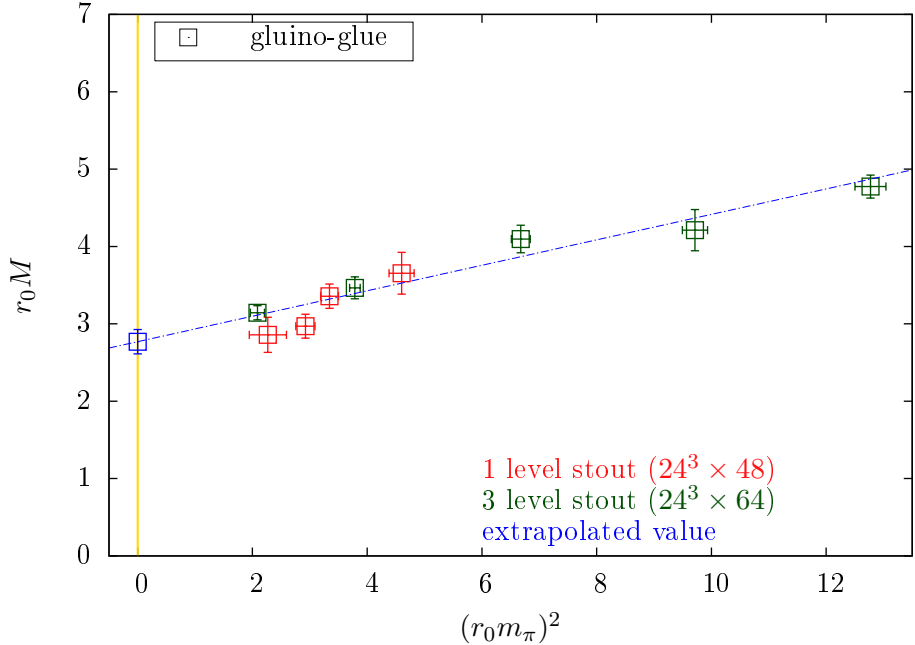}
\caption{The gluino-glue mass as a function of the squared mass of the
adjoint pion in units of the Sommer scale, and the corresponding linear
fit.}
\label{fig:gluinoglue}
\end{center}
\vspace*{-1mm}
\end{figure}

The best accuracy can be obtained for the mass of the gluino-glue particle.
Its correlator has been obtained using a combination of APE and Jacobi
smearing. The gluino-glue mass is shown in Fig.~\ref{fig:gluinoglue} as a
function of the squared adjoint pion mass, together with the extrapolation
to the chiral limit. In all the figures the error bars include both the
statistical and systematic errors.

\begin{figure}[htb]
\begin{center}
\includegraphics{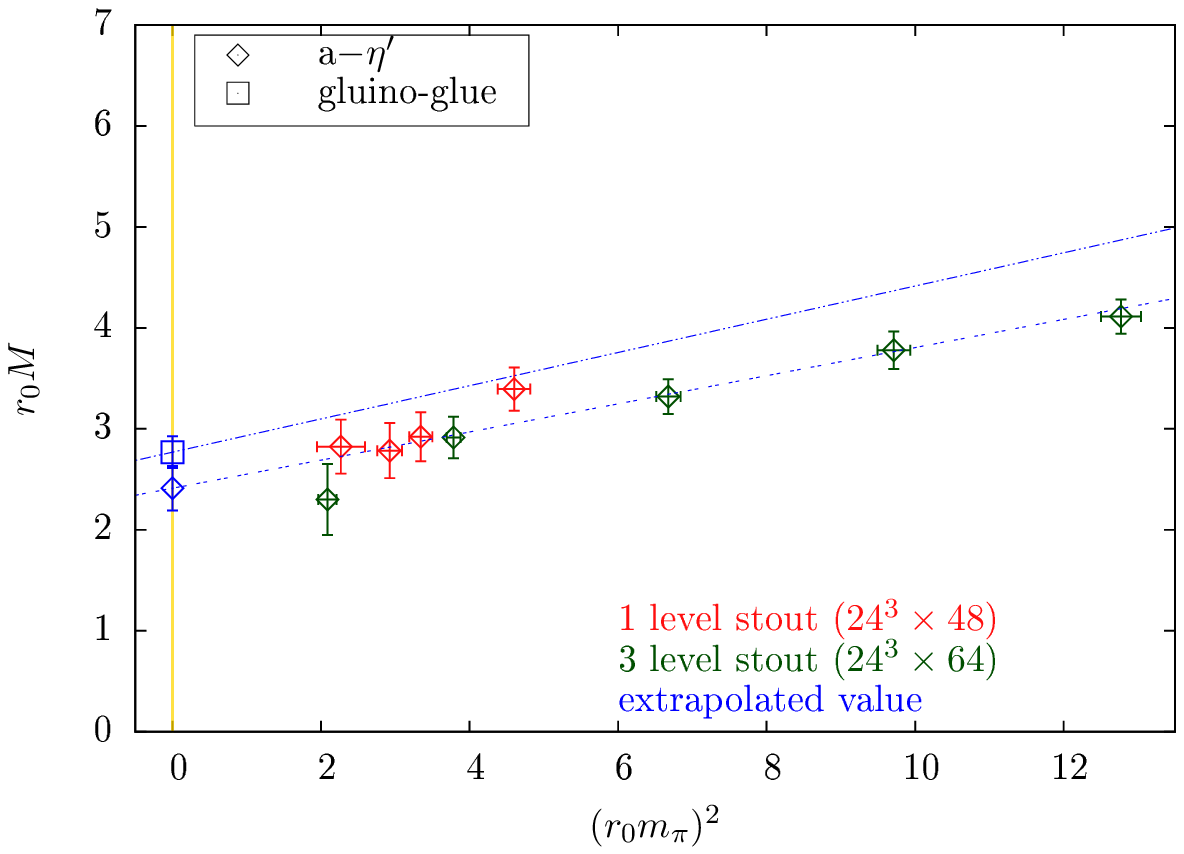}\\[10mm]
\includegraphics{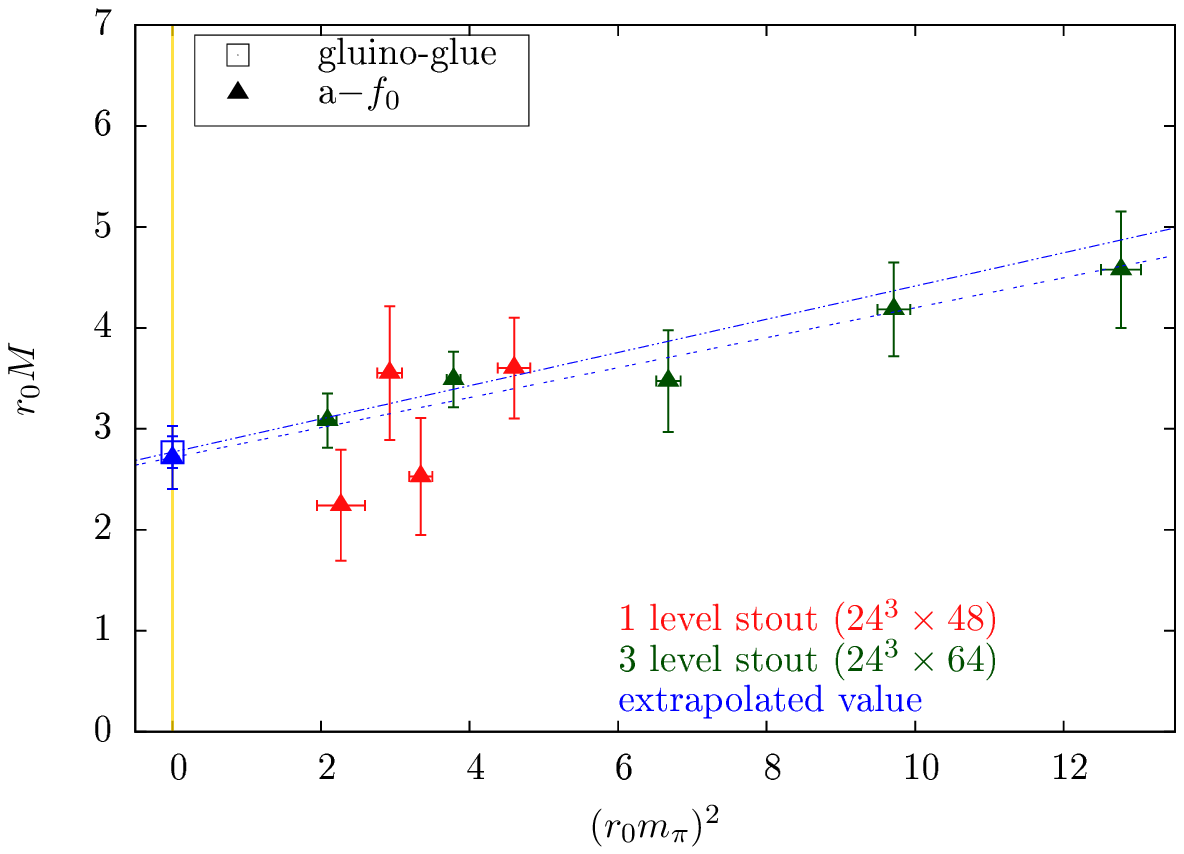}
\caption{The $\aetap$ mass and the $\afn$ mass as functions of the squared
mass of the adjoint pion in units of the Sommer scale, and the corresponding
linear fit. Also shown is the fit for the gluino-glue.}
\label{fig:aetap-afn}
\end{center}
\end{figure}

The correlators for the $\aetap$ and the $\afn$ bosons contain connected and
disconnected contributions. The disconnected parts have been calculated
using the stochastic estimator method \cite{Bali:2009hu}, taking into
account the exact contribution of the 100 lowest eigenmodes of the even-odd
preconditioned Hermitian Wilson-Dirac operator. The disconnected
contributions are especially significant at smaller adjoint pion masses. The
statistical fluctuations in this part are larger than in the connected
contribution and lead to a bad signal-to-noise ratio in the correlators. In
Fig.~\ref{fig:aetap-afn} the masses of the mesons are displayed together
with the extrapolated values at the chiral limit. For comparison, the
figures additionally include the linear fit of the gluino-glue mass.

\begin{figure}[htb]
\begin{center}
\includegraphics{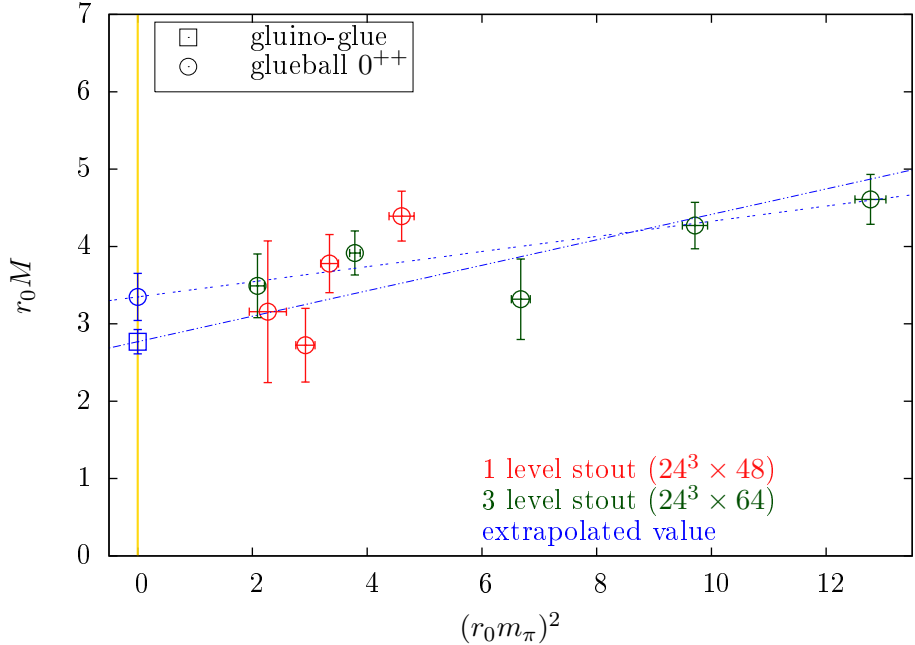}
\caption{The glueball mass as a function of the squared mass of the adjoint
pion in units of the Sommer scale, and the corresponding linear fit. Also
shown is the fit for the gluino-glue.}
\label{fig:gb}
\end{center}
\end{figure}

The operator corresponding to the $0^{++}$ glueball is a combination of
gauge links. The correlation functions are as usual afflicted by large
statistical fluctuations. We are using variational smearing methods to
improve the signal \cite{Luscher:1990ck}. Smearing the underlying gauge
links leads to a basis of different operators with the same quantum numbers.
A combination of these operators providing the best overlap with the
particle state is estimated numerically. From the correlations of the basic
operators the correlation matrix $C(t)$ is obtained. The optimisation
corresponds to the solution of the generalised eigenvalue problem
\begin{equation}
C(t)v=\lambda(t,t_0) C(t_0)v
\end{equation}
for a given fixed $t_0$, in our case $t_0=0$. The resulting eigenvalues
$\lambda(t,0)$ as a function of $t$ are taken as input for a fit procedure
analogous to the case of the mesons $\aetap$ and $\afn$. The quality of the
signal is improved considerably by the variational method. However, the
signal is yet not as good as for the gluino-glue, and the systematic error
estimate of the plateau estimation is taken into account. The results for
the glueball masses are shown in Fig.~\ref{fig:gb}.

\clearpage

Whereas in previous work a significant gap between the masses of the
gluino-glue and its supposed superpartners showed up, the figures presented
here display considerably smaller differences. To make the comparison
quantitative, we confront the present results with those of our earlier work
in \cite{Demmouche:2010sf}, done at a larger lattice spacing. The masses are
summarised in Table \ref{tab:final_res}. They have been converted into units
of MeV using the QCD scale setting $r_0=0.5$ fm.

%%%%%%%%%%%%%%%%%%%%%%%%%%%%%%%%%%%%%%%%%
\begin{table}[htb]
\begin{center}
\begin{tabular}{l|crcr}
\hspace{1mm} $\beta$ & $\aetap$ & $\afn$\hspace{4mm} & $\glg$ & glueball $0^{++}$\hspace*{-3mm}\\
\hline
1.6  & 670(63) &  571(181) & 1386(39) & 721(165)\\
1.75 & 950(87) & 1070(123) & 1091(62) & 1319(120)
\end{tabular}
\caption{Comparison of the bound state masses in units of MeV, extrapolated
to vanishing gluino mass, at the two values of $\beta$. The lattice spacing
at $\beta=1.6$ is between $0.088$ and $0.097$ fm depending on the level of
stout smearing. In the current simulations it is approximately $0.055$ fm
for $l_s=1$, and approximately $0.058$ fm for $l_s=3$. All values are
obtained using the QCD units by setting $r_0=0.5$ fm.}
\label{tab:final_res}
\end{center}
\end{table}
%%%%%%%%%%%%%%%%%%%%%%%%%%%%%%%%%%%%%%%%%

One can observe significant differences between the results at the larger
lattice spacing ($\beta=1.6$) and the current lattice spacing
($\beta=1.75$). The gluino-glue mass gets smaller and the other masses
larger when the lattice spacing is reduced. Taking the statistical and
systematic uncertainties into account, the extrapolations towards vanishing
pion mass of the masses of the fermionic and bosonic states in our present
calculations are consistent with each other. Note that further away from the
chiral limit the difference between fermionic and bosonic masses is expected
to grow.

In order to confirm that the remaining differences between the masses are
finite lattice spacing effects, we are planning to extend the calculations
to a third, even smaller lattice spacing, and to extrapolate the results to
the continuum limit.

%%%%%%%%%%%%%%%%%%%%%%%%%%%%%%%%%%%%%%%%%%%%%%%%%%%%%%%%%%%%%%%%%%%%%%%%
\section{Conclusions}

In this work we present the current results of our simulations of
$\mathcal{N}=1$ supersymmetric Yang-Mills theory on a lattice. Our aim is to
obtain a picture of the bound states of this theory in the nonperturbative
regime. The results of our preparatory study on finite volume effects
\cite{Bergner:2012rv} are complemented with an analysis of the lattice
artifacts. Both, finite volume effects and lattice artifacts increase the
mass gap in the spectrum between the bosonic and the fermionic states. Hence
the scales need to be chosen carefully for a reliable simulation of the
theory.

The parameters of our earlier results \cite{Demmouche:2010sf} have been
chosen according to the experience of the QCD simulations. Our new, more
detailed study has shown that these settings were on the safe side
concerning the finite volume effects. The volume could even be reduced
without a considerable systematic error. The supersymmetry breaking due to
the discretisation effects or lattice artifacts, on the other hand, has been
significant in our earlier results. The difference from the QCD expectations
for the best simulation scales is not unexpected. The volume has to be large
in comparison to the Compton wavelength of the particles under
consideration. In supersymmetric Yang-Mills theory there is no propagating
particle corresponding to the lightest particle of QCD, the pion (the
corresponding operator has been used only to tune the chiral symmetry
restoration). The absence of this light particle induces smaller finite
volume effects.

Our results show that the supersymmetry breaking due to the lattice
artifacts has been significant in our earlier results. With our current
parameters at the smaller lattice spacing, taking the statistical and
systematic uncertainties into account, the extrapolations towards vanishing
gluino mass are consistent with the formation of degenerate supermultiplets.
%This makes it plausible that in the continuum limit there is no spontaneous
%supersymmetry breaking in SYM.
This is an important indication that this supersymmetric theory can be
simulated on the lattice and nontrivial nonperturbative results are consistent
with the theoretical prediction of an absent spontaneous supersymmetry breaking \cite{Witten:1982df}.

%%%%%%%%%%%%%%%%%%%%%%%%%%%%%%%%%%%%%%%%%%%%%%%%%%%%%%%%%%%%%%%%%%%%%%%%
\section*{Acknowledgements}

This project is supported by the German Science Foundation (DFG) under
contract Mu 757/16, and by the John von Neumann Institute for Computing
(NIC) with grants of computing time. Further computing time has been
provided by the compute cluster PALMA of the University of M\"unster.

%%%%%%%%%%%%%%%%%%%%%%%%%%%%%%%%%%


\begin{thebibliography}{99}

\bibitem{Demmouche:2010sf}
  K.~Demmouche, F.~Farchioni, A.~Ferling, I.~Montvay, G.~M\"unster,
  E.~E.~Scholz, J.~Wuilloud,
  %``Simulation of 4d N=1 supersymmetric Yang-Mills theory with Symanzik
  %improved gauge action and stout smearing,''
  Eur.\ Phys.\ J.\ {\bf C 69} (2010) 147
  [\arxiv{1003.2073} [{\tt hep-lat}]].
  %%CITATION = EPHJA,C69,147;%%

\bibitem{Bergner:2011wf}
  G.~Bergner, I.~Montvay, G.~M\"unster, U.~D.~\"Ozugurel, D.~Sandbrink,
  %``Supersymmetric Yang-Mills theory: a step towards the continuum,''
  PoS(Lattice 2011) 055
  [\arxiv{1111.3012} [{\tt hep-lat}]].
  %%CITATION = ARXIV:1111.3012;%%

\bibitem{Amati:1988ft}
  D.~Amati, K.~Konishi, Y.~Meurice, G.~C.~Rossi, G.~Veneziano,
  %``Nonperturbative Aspects in Supersymmetric Gauge Theories,''
  Phys.\ Rept.\ {\bf 162} (1988) 169.
  %%CITATION = PRPLC,162,169;%%
  
\bibitem{Armoni:2009zz}
  A.~Armoni,
  %``QCD and supersymmetry,''
  Nucl.\ Phys.\ Proc.\ Suppl.\ {\bf 195} (2009) 46.
  %%CITATION = NUPHZ,195,46;%%

\bibitem{Bergner:2012rv}
  G.~Bergner, T.~Berheide, I.~Montvay, G.~M\"unster, U.~D.~\"Ozugurel,
  D.~Sandbrink,
  %``The gluino-glue particle and finite size effects in supersymmetric
  %Yang-Mills theory,''
  JHEP {\bf 1209} (2012) 108
  [\arxiv{1206.2341} [{\tt hep-lat}]].
  %%CITATION = ARXIV:1206.2341;%%

\bibitem{Bergner:2009vg}
  G.~Bergner,
  %``Complete supersymmetry on the lattice and a No-Go theorem,''
  JHEP {\bf 1001} (2010) 024
  [\arxiv{0909.4791} [{\tt hep-lat}]].

\bibitem{Curci:1986sm}
  G.~Curci, G.~Veneziano,
  %``Supersymmetry and the Lattice: A Reconciliation?,''
  Nucl.\ Phys.\ {\bf B 292} (1987) 555.

\bibitem{Suzuki:2012pc}
  H.~Suzuki,
  %``Supersymmetry, chiral symmetry and the generalized BRS transformation
  %in lattice formulations of 4D $\mathcal{N}=1$ SYM,''
  Nucl.\ Phys.\ {\bf B 861} (2012) 290
  [\arxiv{1202.2598} [{\tt hep-lat}]].
  %%CITATION = ARXIV:1202.2598;%%

\bibitem{Veneziano:1982ah}
  G.~Veneziano, S.~Yankielowicz,
  %``An Effective Lagrangian for the Pure N=1 Supersymmetric Yang-Mills
  % Theory,''
  Phys.\ Lett.\ {\bf B 113} (1982) 231.

\bibitem{Farrar:1997fn}
  G.~R.~Farrar, G.~Gabadadze, M.~Schwetz,
  %``On the effective action of N=1 supersymmetric Yang-Mills theory,''
  Phys.\ Rev.\ {\bf D 58} (1998) 015009 \\
  {[}\arxiv{hep-th/9711166}\,].

\bibitem{Morningstar:2003gk}
  C.~Morningstar, M.~J.~Peardon,
  %``Analytic smearing of SU(3) link variables in lattice QCD,''
  Phys.\ Rev.\ {\bf D 69} (2004) 054501\\
  {[}\arxiv{hep-lat/0311018}\,].

\bibitem{Stuewe}
H.~Stuewe,
Diploma thesis, Univ.~of M\"unster, forthcoming.

\bibitem{Farchioni:2007dw}
  F.~Farchioni, I.~Montvay, G.~M\"unster, E.~E.~Scholz, T.~Sudmann, J.~Wuilloud,
  %``Hadron masses in QCD with one quark flavour,''
  Eur.\ Phys.\ J.\ {\bf C 52} (2007) 305
  [\arxiv{0706.1131} [{\tt hep-lat}]].

\bibitem{Farchioni:2001wx}
  F.~Farchioni, A.~Feo, T.~Galla, C.~Gebert, R.~Kirchner, I.~Montvay,
  G.~M\"unster, A.~Vladikas,
  %``The Supersymmetric Ward identities on the lattice,''
  Eur.\ Phys.\ J.\ {\bf C 23} (2002) 719
  {[}\arxiv{hep-lat/0111008}\,].

\bibitem{Montvay:2005tj}
  I.~Montvay, E.~Scholz,
  %``Updating algorithms with multi-step stochastic correction,''
  Phys.\ Lett.\ {\bf B 623} (2005) 73
  [\arxiv{hep-lat/0506006}\,].
  
\bibitem{Bergner:2011zp}
  G.~Bergner, J.~Wuilloud,
  %``Acceleration of the Arnoldi method and real eigenvalues of the
  % non-Hermitian Wilson-Dirac operator,''
  Comput.\ Phys.\ Commun.\ {\bf 183} (2012) 299
  [\arxiv{1104.1363} [{\tt hep-lat}]].

\bibitem{Baron:2010th}
  R.~Baron {\it et al.} [European Twisted Mass Collaboration],
  %``Computing K and D meson masses with $N_{f}$ = 2+1+1 twisted mass
  % lattice QCD,''
  Comput.\ Phys.\ Commun.\ {\bf 182} (2011) 299
  [\arxiv{1005.2042} [{\tt hep-lat}]].

\bibitem{Bali:2009hu}
  G.~S.~Bali, S.~Collins, A.~Sch\"afer,
  %``Effective noise reduction techniques for disconnected loops in Lattice
  %QCD,''
  Comput.\ Phys.\ Commun.\ {\bf 181} (2010) 1570
  [\arxiv{0910.3970} [{\tt hep-lat}]].
  
\bibitem{Luscher:1990ck}
  M.~L\"uscher, U.~Wolff,
  %``How To Calculate The Elastic Scattering Matrix In Two-dimensional
  %Quantum Field Theories By Numerical Simulation,''
  Nucl.\ Phys.\ {\bf B 339} (1990) 222.

\bibitem{Witten:1982df}
  E.~Witten,
  %``Constraints on Supersymmetry Breaking,''
  Nucl.\ Phys.\ {\bf B 202} (1982) 253.

\end{thebibliography}
\end{document}